\documentclass[conference]{IEEEtran}
\IEEEoverridecommandlockouts

\usepackage{cite}
\usepackage{amsmath,amssymb,amsfonts}
\usepackage{algorithmic}
\usepackage{graphicx}
\usepackage{textcomp}
\usepackage{xcolor}
\usepackage{multirow}
\usepackage{booktabs} 

\makeatletter
\newcommand{\linebreakand}{%
  \end{@IEEEauthorhalign}
  \hfill\mbox{}\par
  \mbox{}\hfill\begin{@IEEEauthorhalign}
}
\makeatother

\def\BibTeX{{\rm B\kern-.05em{\sc i\kern-.025em b}\kern-.08em
    T\kern-.1667em\lower.7ex\hbox{E}\kern-.125emX}}
\usepackage[most]{tcolorbox}
\newtcolorbox{promptbox}[1][]{ 
  colback=gray!5,              
  colframe=black,
  colbacktitle=gray!5,
  coltitle=black,               
  rounded corners,               
  title,                  
  fonttitle=\bfseries,
  #1                           
}
\begin{document}

\title{SolSearch: An LLM-Driven Framework for Efficient SAT-Solving Code Generation
\thanks{* Equal contribution. This work was supported by National Key Research and Development Program (No.2022YFB3305200), NSFC (No.62161146001, 92370201), STCSM (No.22QB1402100). ({\em{Corresponding author: Xiangfeng Wang and Jianqi Shi}})
}}

\author{\IEEEauthorblockN{Junjie Sheng$^{*}$}
\IEEEauthorblockA{\textit{School of Computer Science and Technology} \\
East China Normal University \\ Shanghai, China \\
jarvis@stu.ecnu.edu.cn}
\and
\IEEEauthorblockN{Yanqiu Lin$^{*}$}
\IEEEauthorblockA{\textit{Software Engineering Institute} \\
East China Normal University\\ Shanghai, China\\
51255902092@stu.ecnu.edu.cn}
\and
\IEEEauthorblockN{Jiehao Wu$^{*}$}
\IEEEauthorblockA{\textit{School of Computer Science and Technology} \\
East China Normal University\\ Shanghai, China \\
51255901123@stu.ecnu.edu.cn}
\linebreakand
\IEEEauthorblockN{Yanhong Huang}
\IEEEauthorblockA{\textit{Software Engineering Institute} \\
East China Normal University\\ Shanghai, China \\
yhhuang@sei.ecnu.edu.cn
}
\and
\IEEEauthorblockN{Jianqi Shi}
\IEEEauthorblockA{\textit{Software Engineering Institute} \\
East China Normal University\\ Shanghai, China \\
jqshi@sei.ecnu.edu.cn}
\and
\IEEEauthorblockN{Min Zhang}
\IEEEauthorblockA{\textit{Software Engineering Institute} \\
East China Normal University\\ Shanghai, China\\
zhangmin@sei.ecnu.edu.cn}
\linebreakand
\IEEEauthorblockN{Xiangfeng Wang}
\IEEEauthorblockA{\textit{School of Computer Science and Technology} \\
East China Normal University\\ Shanghai, China \\
xfwang@cs.ecnu.edu.cn}
}
\maketitle
\begin{abstract}
The Satisfiability (SAT) problem is a core challenge with significant applications in software engineering, including automated testing, configuration management, and program verification. This paper presents SolSearch, a novel framework that harnesses large language models (LLMs) to discover and optimize SAT-solving strategies automatically. Leveraging a curriculum-based, trial-and-error process, SolSearch enables the LLM to iteratively modify and generate SAT solver code, thereby improving solving efficiency and performance.
This automated SAT-solving paradigm has the advantage of being plug-and-play, allowing integration with any SAT solver and accelerating the development or design process of new SAT solvers (new methods). Our preliminary experimental results are encouraging by demonstrating that the LLM-powered paradigm improves state-of-the-art SAT solvers on general SAT benchmarks and significantly enhances the performance of the widely used Z3 solver (11\% on PAR-2 score). These results highlight the potential for using LLM-driven methods to advance solver adaptability and effectiveness in real-world software engineering challenges.
Future research directions are discussed to further refine and validate this approach, offering a promising avenue for integrating AI with traditional software engineering tasks.
\end{abstract}


\section{Introduction}
The Satisfiability (SAT) problem is a foundational concept in computational theory with significant applications across various domains~\cite{implication-color,implication-industry,implication-vertex}, especially in software engineering~\cite{SE-model,SE-theory}. SAT solvers are essential tools for automating tasks such as software testing~\cite{software_testing}, configuration management, and program verification~\cite{verification}. By determining whether a logical formula is satisfiable, they play a crucial role in ensuring software systems' reliability, safety, and performance. Despite their impressive capabilities, how to design a powerful SAT-solver remains an open problem. 

Traditionally, modifying existing SAT solvers or designing new ones relies on rigid, manually tuned techniques tailored to specific problem types~\cite{cadical,kissat}. This approach is labor-intensive, time-consuming, and often lacks generalizability when applied to diverse or novel problem instances. As modern software engineering projects grow in complexity, these limitations will become increasingly burdensome. There is an urgent demand for new automated and adaptive SAT solvers capable of enhancing performance across varied problems.

In recent years, large language models (LLMs) have demonstrated impressive capabilities in mimicking human roles across various tasks, including natural language processing~\cite{LLM-NLP}, coding~\cite{LLM-code}, and problem-solving~\cite{LLM-task}. Beyond these applications, recent studies suggest that LLMs can also assume the role of an expert in innovation, capable of generating novel strategies—for instance, discovering new bin-packing policies~\cite{Funsearch}. This raises an intriguing question: \textit{can LLMs also be harnessed to aid in the discovery of SAT solvers?}

This paper introduces SolSearch, a novel framework that leverages LLMs to automatically discover and optimize SAT-solving strategies, fundamentally transforming SAT solver design. Through a curriculum-based, trial-and-error process, the LLM iteratively modifies portions of the solver’s code to improve efficiency and performance. This dynamic adaptation enables the solver to adjust its approach based on the unique characteristics of each problem, moving beyond static, pre-configured solutions. By integrating LLMs into the optimization process, we transition from rigid, manual adjustments to an automated, flexible methodology capable of adaptively generating novel solving techniques. Extensive experiments show that SolSearch improves state-of-the-art SAT solvers on general benchmarks and significantly boosts the performance of Z3, a widely used solver in software engineering, achieving an 11\% improvement in the PAR-2 score.

\section{Background}
\subsection{The SAT Problem}

\begin{figure*}[t]
\centering
\includegraphics[width=0.9\textwidth]{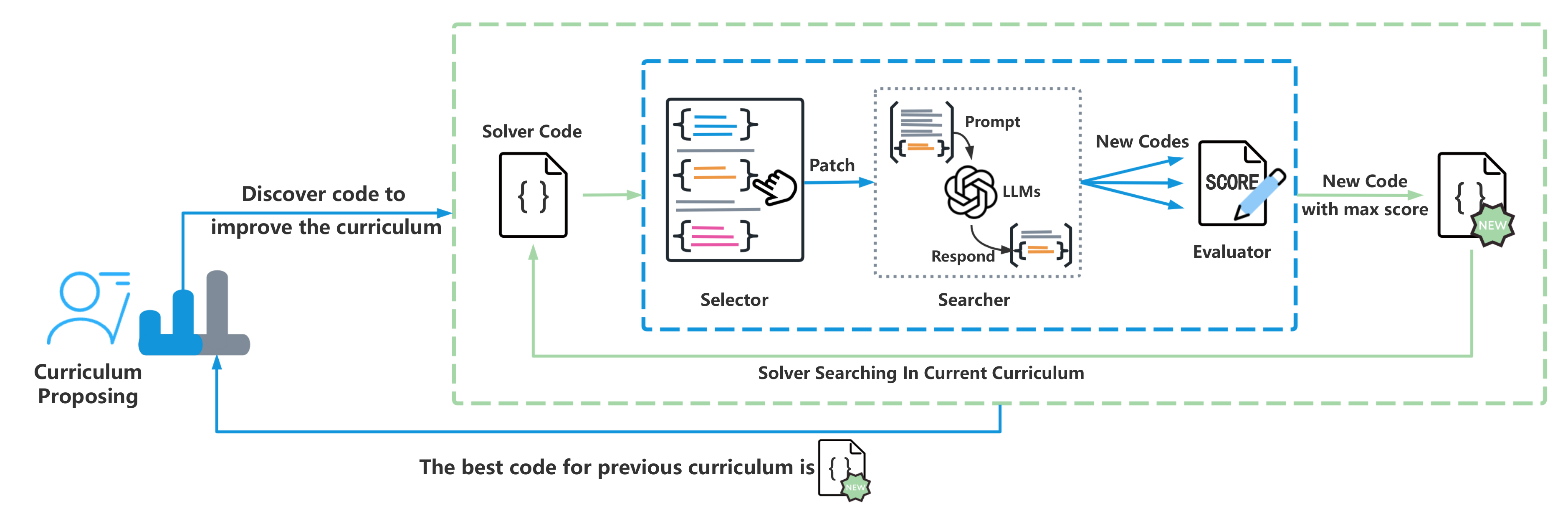}
\caption{Overview of SolSearch. It has two phases: curriculum-proposing and solver-searching. The curriculum-proposing phase creates tasks. The solver-searching phase loops through selector, searcher, and evaluator to improve solvers. This cycle repeats as new tasks are proposed.}
\label{fig:methodology}
\end{figure*}

The satisfiability problem (SAT) determines if there exists an assignment $A$ that makes a given Boolean expression $f$ true. 
In SAT, Boolean expressions are commonly provided in the form of Conjunctive Normal Form (CNF)~\cite{SATproblem}.
Let us consider the set of variables $V = \{v_1, v_2, v_3\}$ and the formulation
\begin{equation}\label{eq:basic-formulation}
    f(V) = (v_1 \wedge \neg v_2) \vee v_3,\nonumber
\end{equation}we can express the above formulation in form of a CNF as
\[
f_0(V) = (v_1 \vee v_3) \wedge (\neg v_2 \vee v_3).
\]
There exists an assignment $A_1(V) = \{1, 1, 1\}$ that results in $f(V) = \text{TRUE}$, demonstrating that $f(V)$ is SATISFIABLE.

The Davis-Putnam-Logemann-Loveland algorithm~\cite{DPLL} is a complete method for solving SAT problems that refine the original Davis-Putnam algorithm by employing a recursive binary tree search to determine satisfiability. Building upon this, the CDCL (Conflict-Driven Clause Learning) algorithm~\cite{CDCL}, which is pivotal in modern SAT solvers~\cite{CDCL-Glucose, CDCL-minisat, CDCL-EVSIDS}, enhances the search process through conflict management and learning mechanisms such as variable selection heuristics and unit propagation. In contrast, Stochastic Local Search (SLS) ~\cite{SLS} utilizes variable flipping strategies to find solutions. SLS-type incomplete methods have proven valuable, especially when complete algorithms are constrained by limited resources.

In both academic and industrial communities, a variety of tools are available, each suited for different purposes. CaDiCaL~\cite{cadical} is an advanced and efficient SAT solver characterized by its concise design, exhibiting outstanding performance in addressing large-scale SAT problems. The Kissat SAT solver~\cite{kissat} builds upon this simplistic design by incorporating optimized structures and algorithms to further improve problem-solving performance. Z3~\cite{Z3}, a theorem prover from Microsoft Research, features a robust API with support for languages like C++ and Python. As an open-source project, Z3 includes a comprehensive SAT solver and offers extensibility.

However, it must be emphasized that designing new SAT solvers under the current mindset is not an easy task. Given the variety of problems and scenarios, developing methods with universal applicability is even more challenging. For instance, in the annual SAT Competition~\cite{satcomp2023,satcomp2024}, the top rankings consistently fluctuate, as shown in Table \ref{tab:satcompresults}. This phenomenon indicates that it is not feasible to develop a universally optimal solver for all SAT problems once and for all.
\begin{table}[htbp]
\centering
\caption{SAT Competition Results for 2022-2024}
\label{tab:satcompresults}
\resizebox{\columnwidth}{!}{
\begin{tabular}{@{}cccc@{}}
\toprule
\textbf{Rank} & \textbf{2022} & \textbf{2023} & \textbf{2024} \\
\midrule
1 & Kissat\_MAB-HyWalk & SBVA Cadical & kissat-sc2024 \\
2 & kissat\_inc & KissatMabProp PrNosym & Kissat\_MAB-DC \\
3 & kissat\_pre & KissatMabProp & hKis-bva \\
\bottomrule
\end{tabular}
}
\end{table}
\subsection{LLMs in Code Generation}

In recent years, the advent of large language models (LLMs) has spurred transformative innovations across diverse fields, particularly in complex tasks like code generation~\cite{code-generation}. For example, the EUREKA project~\cite{Eureka} utilized LLMs to optimize reward function code for complex manipulation tasks. By integrating human feedback for legality and safety, the project achieved results that surpassed expert-crafted solutions, showcasing the ability of LLMs to innovate beyond conventional methods. Similarly, Google DeepMind's FunSearch~\cite{Funsearch} combines a pre-trained LLM with an evaluator to explore solutions within function space. This hybrid approach has yielded groundbreaking results in challenging problems such as the cap set problem and bin packing, outperforming traditional optimization techniques. Evolution of Heuristics (EoH)~\cite{EoH}, merges LLMs with evolutionary computation to automatically design heuristic algorithms. This combination significantly outperforms manual designs created by domain experts, highlighting LLMs’ potential to discover novel solutions that extend the boundaries of existing knowledge.

These advancements indicate that LLMs are not merely tools for automation but are also catalysts for innovation, reshaping how we approach problem-solving and algorithm design. The varied optimization strategies required for SAT solvers across different scenarios underscore the need for such adaptive and innovative capabilities. As the role of LLMs continues to expand, their integration into SAT-solving processes is expected to drive a new paradigm: the enhancement of existing SAT solvers. By leveraging LLMs, we can transcend traditional approaches, enabling the development of SAT solvers that are more flexible, adaptive, and effective across a broad spectrum of applications.

\section{SolSearch}

The proposed SolSearch paradigm is illustrated in Fig.~\ref{fig:methodology} and involves two phases: the curriculum-proposing stage and the solver-searching stage. In the curriculum-proposing stage, optimized curricula are developed to strategically guide the subsequent phase. Following this, the solver-searching stage focuses on searching for the most effective SAT solvers for the curricula at hand. Through repeated curriculum development and solver identification, SolSearch efficiently discovers and tailors high-performing solvers to meet the specific demands of the problem.
Designed for versatility and adaptability, SolSearch functions seamlessly as a plug-and-play module in various software engineering tasks.
\begin{promptbox}[title=Prompt For Solver Searching] 
Optimizing the \texttt{inc\_activity} function within SAT solvers using VSIDS heuristic.

\textbf{Requirements:}
\begin{itemize}
    \item \textbf{Function Name:} \texttt{inc\_activity}
    \item \textbf{Language:} C++
    \item \textbf{Dependencies:} No undefined functions, variables, or external libraries.
\end{itemize}

\textbf{Reference Function:} \texttt{inc\_activity}
\begin{verbatim}
void solver::inc_activity() {
    //example code
}
\end{verbatim}
\begin{itemize}
    \item \textbf{Behavior:} Increases activity by a fixed amount, checks for overflow, and rescales if needed. Manages variable priorities within a priority queue.
\end{itemize}

\textbf{Task:} Develop the optimized \texttt{inc\_activity} function following the guidelines.
\label{prompt}
\end{promptbox}
\subsection{Curriculum-Proposing Stage}

The curriculum-proposing stage of SolSearch plays a crucial role by generating a sequence of tasks, or curricula, to effectively guide the solver-searching process. Tackling highly challenging tasks directly can be difficult; however, by progressing from simpler to more complex tasks (curriculum learning), these tasks become more manageable. This stage is dedicated to designing curricula that are both challenging and optimizable, enabling gradual improvements.

In this phase, SolSearch assesses the characteristics of the current problem and designs a diverse set of training instances that vary in complexity. The primary metric for success is the number of instances solved within a specified timeout limit. Optimization efforts are geared towards maximizing this metric, which encourages solvers to enhance their efficiency.

Following the principles of curriculum learning, the initial timeout is set conservatively as 100 seconds. This strategy challenges the solver with simpler instances initially, promoting rapid improvements by focusing on increasing the number of instances solved within this shorter timeframe. As each optimization cycle concludes, the timeout limit for the subsequent round is increased based on the performance of the newly optimized solver on a test set. 

This approach ensures that solvers face a progressive and systematic challenge, providing consistent optimization opportunities while incrementally raising the difficulty level.

\subsection{Solver-Searching Stage}

The solver-searching stage is dedicated to finding solvers that optimize performance for the specified curriculum. This stage is structured around three main components: the selector, the searcher, and the evaluator, each playing a crucial role in enhancing solver capabilities.

The selector component is charged with deciding which portions of the solver's code should be optimized in the current iteration. This decision process is driven by the solver’s performance metrics on the provided curriculum, pinpointing code segments where significant enhancements are most likely.

The searcher then takes over by using a Large Language Model (LLM) to generate new code implementations for the selected patch. It operates by responding to a detailed prompt that describes the target heuristic functions and outlines specific requirements including dependencies and performance guarantees. These prompts are designed to ensure that the new code not only fulfills basic functional requirements but also integrates effectively with existing components and meets stringent performance standards. Further setting exploration temperatures for LLM, enables the searcher to produce diverse and valid code solutions that adhere to the demanding needs of SAT solver development.

\begin{figure*}[htbp!]
\centering
\includegraphics[width=\textwidth]{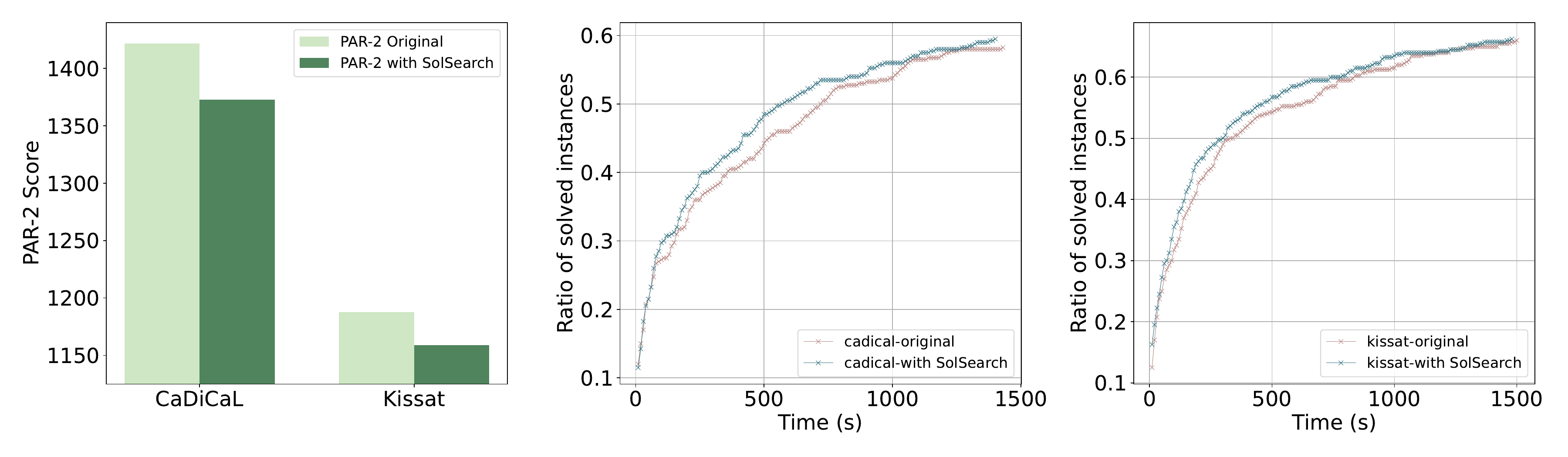}
\caption{Performance comparison of SolSearch on CaDiCaL and Kissat in SAT Competition 2023. Left: SolSearch improves PAR-2 scores for both. Middle and right: SolSearch boosts solved ratios across most timeout settings.}
\label{fig:cadical_kissat}
\end{figure*}

Finally, the evaluator assesses the effectiveness of the newly generated solvers by measuring solution accuracy, computation time, and resource utilization. The top-performing solver from this assessment is chosen and returned to the selector for further refinement in the next cycle. This iterative process continues until the system finds a solver that satisfies the desired performance benchmarks.

Through the coordinated efforts of the selector, searcher, and evaluator, the solver-searching stage methodically explores and implements potential improvements, leading to the development of an optimized solver that is ideally suited for the task at hand. This systematic process ensures continuous enhancement and fine-tuning of solver performance.

\section{Experimental Exploration}

To evaluate the effectiveness of SolSearch, we conducted experiments to answer two main research questions: 
1) Can SolSearch improve state-of-the-art (SOTA) SAT solvers?
2) Can SolSearch be integrated into Z3 in specific tasks?

\paragraph{Experimental Setup}
We conducted our experiments using a dataset from the SAT Competition $2023$, which consists of $400$ benchmark SAT instances representing a diverse range of difficulty levels. These instances were used as the test set to assess the capabilities of SolSearch comprehensively.
The baseline solvers used in our experiments included established SAT solvers such as CaDiCaL~\cite{cadical} and Kissat-sc2024~\cite{kissat} from academic research, as well as the theorem prover Z3~\cite{Z3} from industrial practice. We use deepseek-coder as the LLM in SolSearch. Two primary metrics were used to evaluate performance: the number of solved instances within a given timeout limit and the PAR2 score~\cite{PAR-2}, which calculates the average runtime for solved instances and assigns twice the timeout limit for unsolved instances.

\paragraph{Improving SOTA SAT Solvers}
To address the first research question, we applied SolSearch to CaDiCaL and Kissat-sc2024, and analyzed its impact on their performance in SAT competition 2023. The performance results of CaDiCaL and Kissat before and after optimization are detailed in Fig.~\ref{fig:cadical_kissat}.
The results showed that SolSearch improved the performance of both solvers, solving more instances within the timeout and reducing PAR2 scores.  The optimized versions consistently outperformed the original versions over time. This validates the effectiveness of SolSearch in enhancing solver capabilities.

\paragraph{Enhancing Z3 Solver}
We integrated SolSearch with Z3 to evaluate its impact. Taking the Knights Tour problem (40 instances) proposed in SAT Competition 2024 as the case, Z3 can only solve 17 instances. 
Table \ref{tab:kngihts_tour} shows that Z3 with SolSearch solves more instances within the timeout and obtains better PAR2. 
These findings suggest that SolSearch can be used as a plug-and-play module to enhance solvers like Z3, offering a practical solution for industrial applications.

\begin{table}[htbp]
\centering
\caption{Comparison of Z3 and Z3 w/ SolSearch in Knights Tour. }
\label{tab:kngihts_tour}
\begin{tabular}{lcc}
\toprule
         & Z3 & Z3 w/ SolSearch  \\
\midrule
Solved Ratio in t $\leq$ 100s     & 7.5\%                  & \textbf{10.0\% }                  \\
Solved Ratio in t $\leq$ 300s     & 35.0\%                  & \textbf{42.5\%}                  \\
Solved Ratio in t $\leq$ 500s     & 42.5\%                  & \textbf{52.5\%}                 \\
\midrule
PAR-2            & 660.09              & \textbf{583.52 }             \\
\bottomrule
\end{tabular}
\end{table}
\section{Future Plans}
Building on the promising results in LLM-driven rule extraction, we have outlined our plans to improve our paper. 

\textbf{Insight into Solver Design}: Analyzing the solvers generated by our framework offers opportunities to gain deeper insights into SAT solver design principles. Studying the novel functions and strategies identified by the LLM can inspire innovative directions for SAT solver development.

\textbf{Expanding to Other Tasks}: There is significant potential to extend this framework to tackle other complex tasks, such as Max-SAT~\cite{MaxSAT}, and QBF~\cite{QBF}. By testing our LLM-driven approach on these additional tasks, we can explore the generalizability and adaptability of the framework, opening up new avenues for formal verification and problem-solving across diverse domains.

\textbf{Evaluating LLM Impact}: A deeper analysis of how LLM selection (such as model size, training data, and specialization) influences solver performance and output quality would be valuable for optimizing SAT-solving strategies.



\newpage

\bibliographystyle{IEEEtran}
\bibliography{IEEEexample}

\end{document}